\documentclass[aps,prl,reprint,twocolumn,superscriptaddress,english,longbibliography,floatfix]{revtex4-2}
\usepackage{amssymb}
\usepackage{graphicx}
\usepackage{amsmath}
\usepackage{epsfig}
\usepackage[latin9]{inputenc}
\usepackage{array}
\usepackage{multirow}
\usepackage{color}
\usepackage{esint}
\usepackage{bm}
\usepackage{bbm}
\usepackage{epstopdf}
\usepackage{mathtools}
\usepackage{soul}
\usepackage[dvipsnames]{xcolor}
\usepackage{tikz}
\usepackage[bookmarks=true,pdftex,colorlinks=true,urlcolor=blue,linkcolor=black,citecolor=blue,breaklinks=true,hypertexnames=false]{hyperref}
\usepackage{multibib} 
\usepackage{physics}

\begin{document}
\title{Realizing a spatially correlated lattice interferometer}

\author{Peng Peng}
\affiliation{State Key Laboratory of Advanced Optical Communication System and Network, School of Electronics, Peking University, Beijing 100871, China}

\author{Dekai Mao}
\affiliation{State Key Laboratory of Advanced Optical Communication System and Network, School of Electronics, Peking University, Beijing 100871, China}

\author{Yi Liang}
\affiliation{State Key Laboratory of Advanced Optical Communication System and Network, School of Electronics, Peking University, Beijing 100871, China}

\author{Guoling Yin}
\affiliation{State Key Laboratory of Advanced Optical Communication System and Network, School of Electronics, Peking University, Beijing 100871, China}

\author{Hongmian Shui}
\affiliation{State Key Laboratory of Advanced Optical Communication System and Network, School of Electronics, Peking University, Beijing 100871, China}

\author{Bo Song}\email{bsong@pku.edu.cn}
\affiliation{State Key Laboratory for Mesoscopic Physics and Frontiers Science Center for Nano-optoelectronics, School of Physics, Peking University, Beijing 100871, China}

\author{Xiaoji Zhou}\email{xjzhou@pku.edu.cn}
\affiliation{State Key Laboratory of Advanced Optical Communication System and Network, School of Electronics, Peking University, Beijing 100871, China}
\affiliation{Institute of Carbon-based Thin Film Electronics, Peking University, Shanxi, Taiyuan 030012, China}

\date{\today}

\begin{abstract}

Atom interferometers provide a powerful tool for measuring physical constants and testifying fundamental physics with unprecedented precision. Conventional atom interferometry focuses on the phase difference between two paths and utilizes matter waves with fixed coherence.
Here, we report on realizing a Ramsey-Bord\'e interferometer of coherent matter waves dressed by a moving optical lattice in the gravity direction, and explore the resulting interference along multiple paths with tunable coherence. We investigate spatial correlations of atoms both within the lattice and between two arms by interferometry, and observe the emerging multiple interference peaks owing to the long-range coherence nature of the Bose-Einstein condensate. Our findings agree well with theoretical simulations, paving the way for high-precision interferometry with ultracold atoms.
         
\end{abstract}
    
\maketitle

Atom interferometry is a high-precision method for probing physical quantities and testifying fundamental physics, ranging from inertial
sensing, gravitational wave detection, ultralight dark matter and dark energy search, to testing the Standard Model and quantum mechanics in new regimes~\cite{cronin2009optics,dimopoulos2008atomic,dimopoulos2007testing,muller2008atom,parker2018measurement,asenbaum2020atom}. Recent advances in cold atoms enable atom interferometers as quantum sensors for real-world applications, including gravitational wave detection using long baseline interferometers~\cite{bongs2019taking, abe2021matter,badurina2020aion,zhan2020zaiga}. 
However, traditional matter-wave interferometers in a gravitational field use freely falling atoms~\cite{peters1999measurement,dickerson2013multiaxis} and their sensitivity is limited by the interrogation time of falling atoms. More recent implementation of an optical lattice substantially increases this time by holding atoms in an optical lattice, leading to a considerable enhancement in detection sensitivity~\cite{xu2019probing}. On the other hand, typical atom interferometer experiments use sources that are lack of long-range coherence, and are limited by the classical precision limit~\cite{peters1999measurement,dickerson2013multiaxis,xu2019probing}. Instead, coherent matterwave sources such as Bose-Einstein condensate (BEC) have quantum wave-like behaviours and long-range phase coherence~\cite{andrews1997observation,bloch2000measurement,GUO20222291}, providing more opportunities for quantum-enhanced measurements~\cite{giovannetti2004quantum,esteve2008squeezing,gross2010nonlinear,karcher2018improving}. Coherent matter waves dressed by optical lattices in a gravitational field will bring novel and rich phenomena to atom interferometers.

Here we report on a spatially correlated interferometer with coherence-tunable matter waves dressed by an optical lattice in the gravity direction, and investigate the effect of coherence on both the interference of atoms in two arms and the interference 
 of atoms within the lattice, termed ``intraference''. The long-range coherence plays a vital role in precision measurements, e.g. atom interferometers using a BEC can be robust against systematic noise and imperfections including the imperfect Raman pluses~\cite{hardman2014role}. We have observed the emerging multipath interference due to the long-range coherence feature of the BEC, and studied the amplitude change and position shift of the interference contrast.
Intraference and interference between two arms together give rise to a multipath meshgrid interference. 

Moreover, using the interference effect, we have precisely measured the correlation of atoms in the optical lattice. So far, the spatial distribution of atoms has been measured with sub-lattice resolutions by superresolution or bilayer lattice techniques~\cite{mcdonald2019superresolution,subhankar2019nanoscale,du2024atomic}, whereas the precise measurement of spatial correlations remains difficult. In contrast to the conventional time-of-flight method only detecting the overall coherence of the system, which is limited by the expansion time and the initial cloud size~\cite{gerbier2008expansion}, interferometry here provides more accurate and detailed spatial correlations. 

\begin{figure*}[htbp]
\centering
\includegraphics[width=0.9\textwidth]{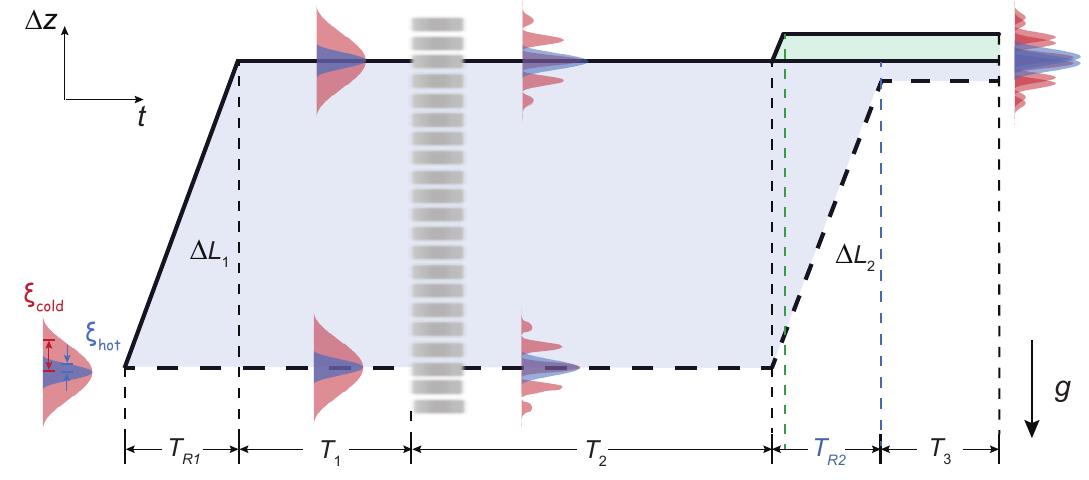}
\caption{\textbf{Space-time representation of the atom interferometer.}
The atom interferometer is based on a Ramsey-Bord\'e interferometer dressed by a moving optical lattice. Two arms of the interferometer (solid and dashed lines indicate the upper arm and the lower arm respectively) are split by a pair of $\pi/2$ Raman pulses separated by $T_{R1}$, and recombined by another pair separated by $T_{R2}$.
Two arms are separated by $\Delta L_1=2v_{\rm {rec}}T_{R1}$  after the first pair, and atoms spaced by $\Delta L_2=2v_{\rm {rec}}T_{R2}$ interfere after the second pair.
In between two pairs, atoms freely fall for $T_1$ before being loaded into a moving lattice (gray), followed by lattice unloading and an extra $T_2$ free fall. Eventually, the interference pattern is detected after the last free fall for $T_3$.
Red and blue indicate cold and hot atoms respectively, and areas show the correlation with a coherence length of $\xi$.
Horizontal and diagonal lines represent $|F=1\rangle$ and $|F=2\rangle$ states respectively. 
The light blue areas are the interference between the upper and lower arms, and the light green area is the interference within the optical lattice, which is termed ``intraference''. The blue and green dashed lines represent the final Raman $\pi/2$ pulse for those two cases.
\label{fig:Schematic}}
\end{figure*}

\section{Atom Interferometer}
Our atom interferometer is built upon a Ramsey-Bord\'e interferometer dressed by a moving optical lattice. The experiment begins with a BEC or a gas of around $10^5$ $^{87}$Rb in the $|F=1, m_F=0\rangle$ state. Atoms are prepared with different temperatures by controlling the final depth of a crossed optical dipole trap (see Methods for details). Then, atoms are suddenly released from the trap, free fall along the gravity direction and enter the interferometer.

Fig.~\ref{fig:Schematic} shows the experimental protocol of the atom interferometer. Atoms are first split by a pair of $\pi/2$ Raman pulses. The duration between the first pair Raman pulses is $T_{R1}$. Two split atomic clouds with the same velocity are separated along the gravity direction with a distance $\Delta L_1=2v_{\rm {rec}}T_{R1}$, where the atomic recoil velocity is $v_{\rm {rec}}=\hbar k_R/m_{\rm Rb}$ with the reduced Planck constant $\hbar$, the wave number of the Raman beam $k_R$, and the mass of the $^{87}$Rb atom $m_{\rm Rb}$. 

Atoms are loaded into a moving optical lattice in $1\,$ ms. The laser light of the optical lattice is $45\,$GHz blue-detuned from the $|F=1\rangle \rightarrow |F'=0\rangle$ transition. Atoms remain stationary with respect to the optical lattice frame by tuning the frequency difference between the two lattice beams. Then atoms are held in the lattice for extra $6\,$ms, during which two atomic clouds undergo 59 Bloch oscillations, i.e. $118 \hbar k_L$  momentum transfer along the opposite direction of the gravity. $k_L=2\pi/\lambda_L$ is the wave number of the lattice beam with the lattice wavelength $\lambda_L$. After atoms are unloaded from the lattice in 1ms, they free fall for $T_2$  before the second pair of Raman pulses are shined. The second pair recombines two arms and the interval between pulses $T_{R2}$ determines the shift between two clouds,  $\Delta L_2=2v_{\rm {rec}}T_{R2}$. The atomic clouds are detected after the last free fall for $T_3$. In the experiment, the splitting and recombination of atomic clouds can be precisely controlled by $T_{R1}$ and $T_{R2}$. For example, when $\Delta L_1 = 30d$ and $\Delta L_2\sim d$ with the lattice spacing $d=\lambda_L/2$, only the intraference within the optical lattice can happen, whereas for $\Delta L_2\sim \Delta L_1$, the upper and lower arms interfere in space.  

\begin{figure*}[htbp]
\centering
\includegraphics[width=0.9\textwidth]{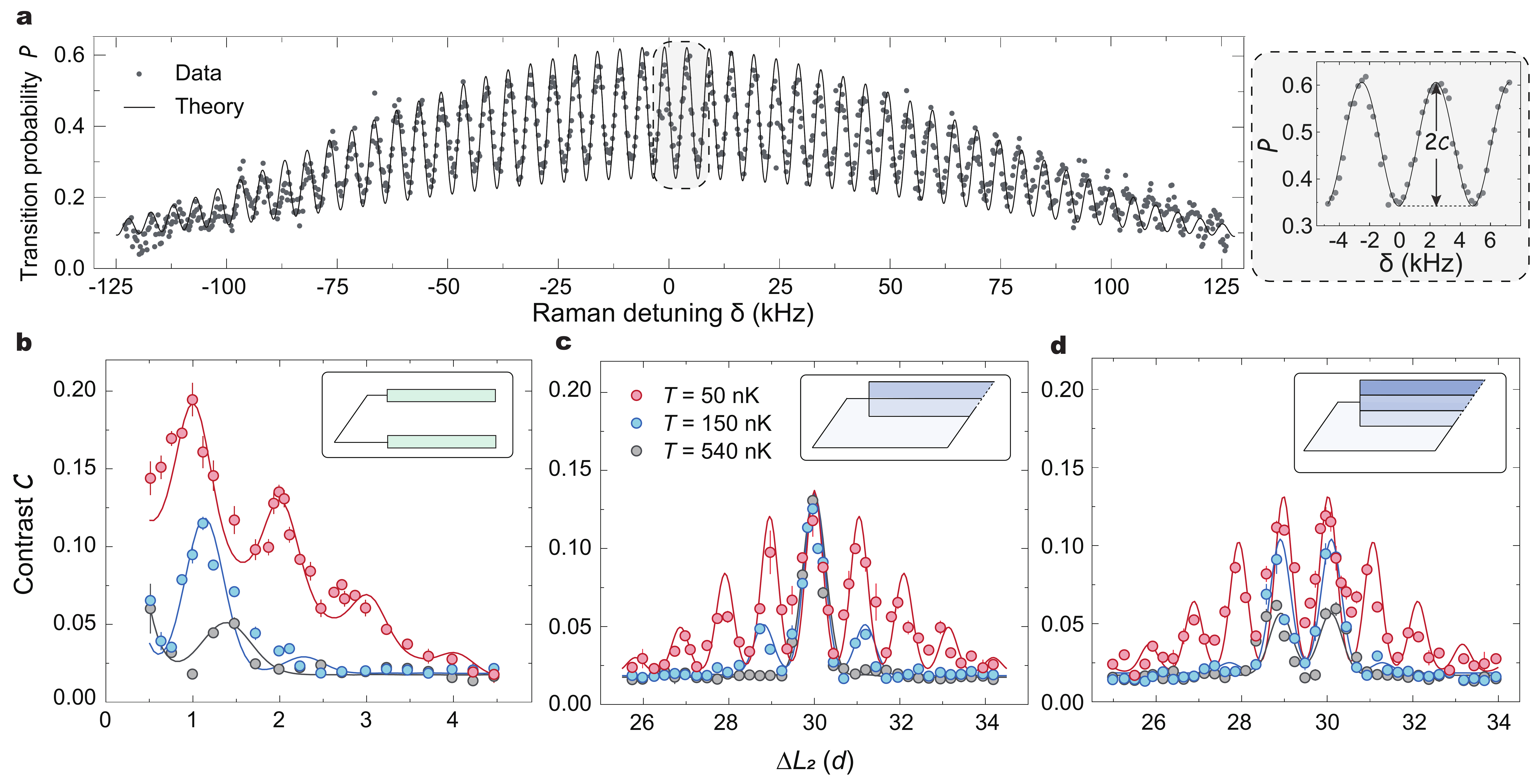}
\caption{\textbf{Interference fringes and contrast.} (\textbf{a}) Representative Ramsey-Bord\'e fringes with atoms dressed by a moving optical lattice. The transition probability $\mathcal{P}$ oscillates with the Raman detuning $\delta$, from which the contrast $\mathcal{C}$ is extracted.
(\textbf{b-d}) Contrast of interference fringes as a function of $\Delta L_2$ for a fixed $\Delta L_1=30d$. Data points are the average over three repetitions, and solid lines are simulations. Noticeable peak shifts away from integer lattice spacings are governed by the coherence length and the momentum distribution of atoms. Coherence lengths used in the simulation are $1.0d$, $1.7d$ and $3.8d$ for $T=\,50\,$nK, $150\,$nK and $540\,$nK respectively.
(\textbf{b}) For $\Delta L_1\gg\Delta L_2$, two arms are not recombined and only the intraference can exist. For $\Delta L_2=n d$, atoms separated by $n d$ interfere. With decreasing the temperature of atoms, the coherence length becomes longer, resulting in a higher contrast of the interference and more visible interference peaks.
(\textbf{c}) For $\Delta L_1\approx\Delta L_2$, two arms are recombined within the coherence length. When the lattice loading is an integer lattice loading, an interference peak occurs at $L_2=L_1$. Similarly, more peaks become visible for lower temperature.
(\textbf{d}) By contrast, for the half-integer lattice loading, with $\Delta L_1=29.5d$, a double-peak pattern can occur even for hot atoms. More peaks become visible for colder atoms. Insets in (b-d) indicate the type of interference, and green for intraference only and blue for interference.\label{fig:Ramsey}}
\end{figure*}

\section{Interference and Intraference}

Fig.~\ref{fig:Ramsey} shows a representative Ramsey-Bord\'e interference pattern and extracted contrasts for different types of lattice loading. 
The transition probability $\mathcal{P}=N_2/(N_1+N_2)$ is measured as a function of the two-photon detuning of the second pair of Raman pulses $\delta$, reflecting the phase change of the interference, where $N_1$ and $N_2$ are the numbers of atoms in $|F=1\rangle$ and $|F=2\rangle$ states respectively. The contrast of the oscillatory transition probability, $\mathcal{C}$ reflects the coherence of the system. Ramsey-Bord\'e fringes have a background around the resonance, in agreement with our theoretical simulation (solid lines; see Methods for details). We measure the contrast as a function of $\Delta L_2$. We first explore the interference only involving atoms in different lattice sites within the same arm, i.e. ``intraference'' in Fig.~\ref{fig:Ramsey}b.
Atoms in different sites evolve differently in phase due to the gravitational potential, yielding different interference patterns. 
We study the impact of coherence on the fringes by varying the temperature of atoms. When the temperature is high ($T=540\,$nK), the contrast is almost zero, indicating a very short coherence length. With decreasing the temperature ($T=150\,$nK), the contrast increases and the emergence of the peak at $\Delta L_2/d\approx 1$ indicates the nearest neighbour interference. When atoms are further cooled down ($T=50\,$nK), atoms enter the stage of the BEC and more peaks become visible when $\Delta L_2$ is around an integer number of the lattice spacing as $\Delta L_2/d\approx n$. 
The $n$th peak indicates the interference between atoms spaced by $n d$, showing the range of the coherence. The behaviour of multi-peak interference manifests the long-range coherence feature of the BEC.

On the other hand, the upper and lower clouds can interfere by varying the time between Raman pulses. The contrast of interference is observed within the coherence length of atoms for $\Delta L_2=\Delta L_1$. Atoms in different lattice sites can also form a series of sub-interferometer within the coherence, resulting in a multiple path interference. It is worth noting that the multiple path is an alternative way to improving the precision of the interferometer sensitivity by effectively narrowing down the width of fringes~\cite{weihs1996all,weitz1996multiple,hinderthur1999time,hadzibabic2004interference,robert2010quantum,petrovic2013multi}. We first study the condition when  $\Delta L_1$ is an integer number of the lattice spacing. Similarly, for high temperatures, the only visible peak is at $\Delta L_2=\Delta L_1$, which reflects the fast decay of correlation in space. With decreasing the temperature, higher order peaks up to the fourth order can be observed at $T=50\,$nK in Fig.~\ref{fig:Ramsey}c, manifesting the long-range coherence of the BEC. 
Moreover, we measure the atom interference for the half-integer lattice loading, $\Delta L_1 = (n+1/2)d$.
In contrast to the integer lattice loading, half-integer lattice loading can have a double-peak pattern, when the two waves have different phases and thus destruct in the middle of two peaks.
A multiple-peak feature can appear when the matter wave has a longer coherence length, e.g. $T=50\,$nK in Fig.~\ref{fig:Ramsey}d.

\begin{figure*}[htbp]
\centering
\includegraphics[width=\textwidth]{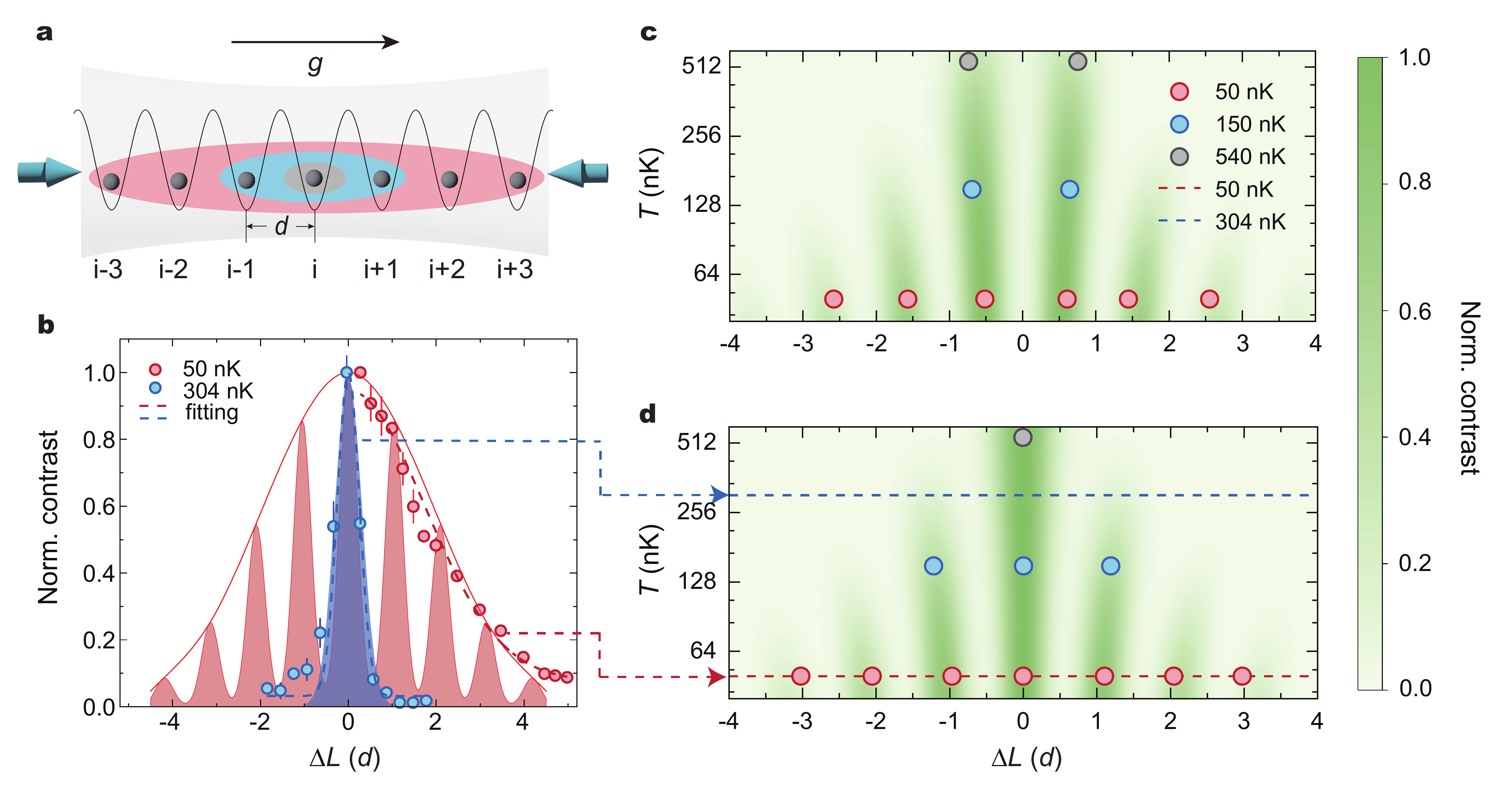}
\caption{\textbf{Temperature-dependent shift of the contrast peak.} 
(\textbf{a}) Schematic shows the correlations of atoms between different sites in an optical lattice for different temperatures. Atoms with lower temperature (red) have a longer coherence length $\xi$.
(\textbf{b}) Normalized correlation of atoms as a function of the distance, $\Delta L=\Delta L_2-\Delta L_1$ for hot atoms (blue, with a temperature of $T=304\,$nK) and cold atoms (red, $T=50\,$nK). The filled areas are calculated correlations of atoms modified by an optical lattice, and its envelope (solid lines) can reflect the correlation of atoms without lattices. The open circles and dashed lines are normalized contrast of fringes without lattices and the fitting of the contrast using Eq.~\ref{coherence length} for different temperatures.
(\textbf{c, d}) Contrast of interference fringes as a function of $\Delta L$ and temperature. They are half-integer lattice loading with $\Delta L_1 = 29.5d$, and integer lattice loading with $\Delta L_1 = 30d$ respectively. The green background is a theoretical simulation and solid circles are the peak positions extracted from Fig.~\ref{fig:Ramsey}(c, d).
\label{figS:2DTemp}}
\end{figure*} 

\section{Shift of Interference peaks}

Notably, the temperature of atoms not only affects the amplitude of interference peaks but also modifies their positions. When the temperature is increased, peaks first tend to shift outwards away from the original lattice in the intraference in Fig.~\ref{fig:Ramsey}b. This effect has also been observed in the interference for both integer and half-integer lattice loading (Fig.~\ref{fig:Ramsey}c,d). We extract the positions of interference peaks by fitting a multiple Gaussian curve to data (see Methods for the detailed analysis).  

Fig.~\ref{figS:2DTemp} shows the temperature-dependent shift of the peak from the lattice site. The resulting shift is attributed by the combined effect of the coherence and the spread of the atomic cloud. The colder atomic cloud has a longer coherence length (red in Fig.~\ref{figS:2DTemp}a) and a narrower momentum distribution. We further confirm the coherence is mainly determined by the temperature via comparing it with the measured contrast without the lattice (open circles in Fig.~\ref{figS:2DTemp}b), which is an envelope of the simulated correlation (red shaded area). In contrast, the hotter atoms (blue) have a shorter coherence length and its correlation function quickly decays with increasing the distance. Extracted peaks' positions agree well with the simulation (green) for both half-integer and integer lattice loading in Fig.~\ref{figS:2DTemp}(c,d).

\section{Time-Evolving Interference}

\begin{figure} [h!]
\centering
\includegraphics[width=0.48\textwidth]{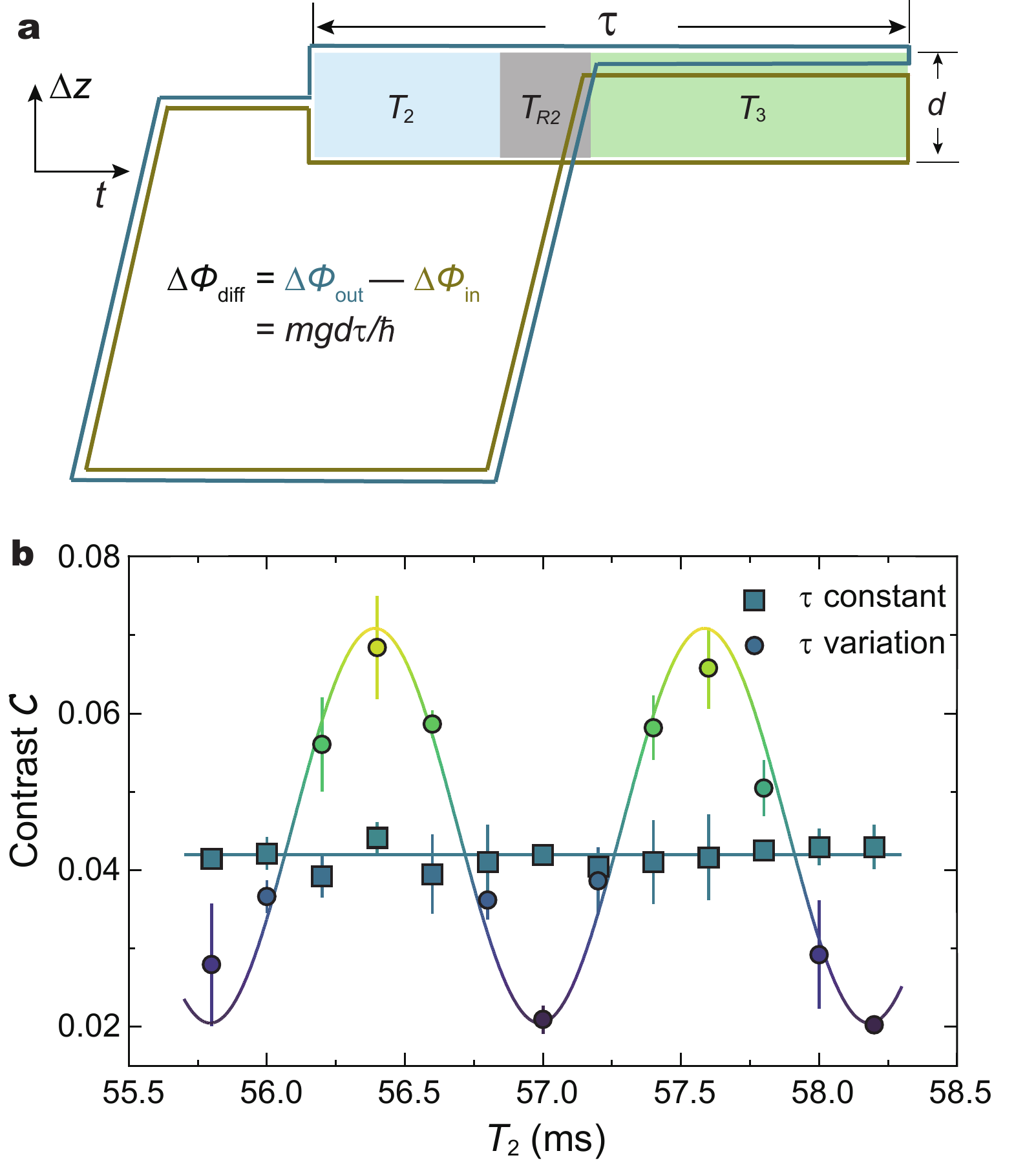}
\caption{\textbf{Time-evolving interference for half-integer lattice loading} (\textbf{a}) The phase evolution of the atom interferometer for half-lattice loading. Two sub-paths separated by a lattice spacing $d$ of the upper arm can both interfere with the lower arm, leading to the phase of the interference evolving with time. The spacing can be tuned by controlling the difference between $\Delta L_1$ and $\Delta L_2$. The shaded areas represent the differential interference phase between two paths, which is proportional to the space-time area $S=\tau d$ with the time between the lattice loading and the detection, $\tau$. The illustration shows the relationship between the phase difference ${\Delta \varPhi_{\rm diff}=\Delta \varPhi_{\rm out}-\Delta \varPhi_{\rm in}}$ and the durations $T_2$, $T_3$. (\textbf{b}) The contrast of the fringes is plotted as a function of $T_2$ for the half-integer lattice loading using parameters of $(\Delta L_1, \Delta L_2)= (29.5d, 29.5d)$ at $T=150\,$nK for fixed (squares) or varying (circles) $\tau$. The phase of the interference is reflected by the contrast of the fringes. It remains almost constant (squares) for a fixed $\tau$, whereas it oscillates with $\tau$ in a period of $1.2$ms (circles). They together show that the interference not only relies on the time before the recombination of two arms, ($T_2$+$T_{R2}$) but also when the interference pattern is detected $T_3$.
\label{fig:Diff}}
\end{figure}

For a typical atom interferometer, the interference pattern is fixed and independent on when the detection is conducted. However, we have observed the time-evolution of the interference pattern for the half-integer lattice loading. We investigate this phenomenon by effectively varying the time between the last $\pi/2$ pulse and the detection, and highlight the time of the detection is also important for atom interferometers. The upper and lower clouds dressed by the optical lattice are mismatched by half a lattice spacing in the gravity direction when two clouds meet. They evolve asynchronously due to different gravitational potentials, resulting in an evolving interference pattern. 

Fig.~\ref{fig:Diff}a shows the accumulated phase (in the shaded rectangular) between the splitting and recombining steps $(T_2+T_{R2})$ in light blue and grey areas respectively, and the time before the detection $T_3$ in the light green area. The phases of two adjacent atomic interferometers are $\Delta \varPhi_{\rm out}$ and $\Delta \varPhi_{\rm in}$ respectively. The differential phase ${\Delta \varPhi_{\rm diff}=\Delta \varPhi_{\rm out}-\Delta \varPhi_{\rm in}}=mgd\tau/\hbar$ reflects the actual phase between two interferometers. The phase of the interference, reflected by the contrast, is independent on when the last pair of Raman pulses as long as the time between the lattice loading and detection is fixed. The contrast oscillates with the total time $\tau=T_2+T_{R2}+T_3$ (open circles) and the color indicating the phase, whereas the phase remains constant (open squares) for a fixed $\tau$ in Fig.~\ref{fig:Diff}b. Here we use atoms with a temperature of $T=150\,$nK. This multipath interference is different from conventional atom interferometers, providing a more precise tool for probing non-uniform field. A similar effect with cold atoms in an optical lattice has been observed by Ref.~\cite{xu2019probing}. Our findings show that the time of the detection is critical for the atom interferometer when the matter wave is coherent in particular.

\paragraph{\textbf{Conclusions.}} We realize a novel Ramsey-Bord\'e interferometer with  coherence-tunable matter waves dressed by a moving optical lattice. We highlight the emerging multiple interference peaks due to the long-range coherence of the BEC, and study the amplitude and shift of those peaks at different temperatures. We also demonstrate precise measurements of spatial correlations for different temperatures and lattice loading using interferometry. Our findings bridge the gap between coherent matter waves and atom interferometry, opening the door for high-precision interferometry with ultracold atoms.

\paragraph{\textbf{Acknowledgments.}}   This work was supported by National Key Research and Development Program of China (Grants No. 2021YFA0718300) and the National Natural Science Foundation of China (Grants No. 92365208, No. 11934002). B.S. acknowledges support from The Fundamental Research Funds for the Central Universities, Peking University and Natural Science Foundation of China (Grant No. 12374242).  

\bibliography{AI.bib}
\subsection{Methods}

\setcounter{equation}{0}
\setcounter{figure}{0}
\setcounter{table}{0}
\renewcommand{\thefigure}{S\arabic{figure}}
\renewcommand{\theequation}{S\arabic{equation}}
\renewcommand{\theHfigure}{S\thefigure}

\paragraph{\textbf{Experimental setup and sequence.}} Our experiments begin with preparing cold atoms or a Bose-Einstein condensate (BEC) of around $1.5\times 10^{5}$ $^{87}$Rb atoms in the $\ket{F=1, m_F=0}$ state. 
The temperature of atoms is in a range from $T=50\,$nK to $540\,$nK by adjusting the final power of the crossed optical dipole trap (CODT) at the end of evaporative cooling.

Then atoms are split by a pair of two counter-propagating Raman beams and have different momenta, leading to different atomic trajectories. The Raman beam is $825\,$MHz red-detuned from the atomic resonance. The two-photon Raman transition couples $\ket{F=1}$ and $\ket{F=2}$ states, resulting in a momenta transfer, $\hbar k_{\rm R}/m_{\rm Rb}=11.8\, {\rm mm/s}$ where $k_{\rm R}$ is the wave number of the Raman beam. Four $\pi/2$ Raman pulses are used to split and recombine atomic clouds for creating the atom interferometer.

\begin{figure}[h]
\centering
\includegraphics[width=0.35\textwidth]{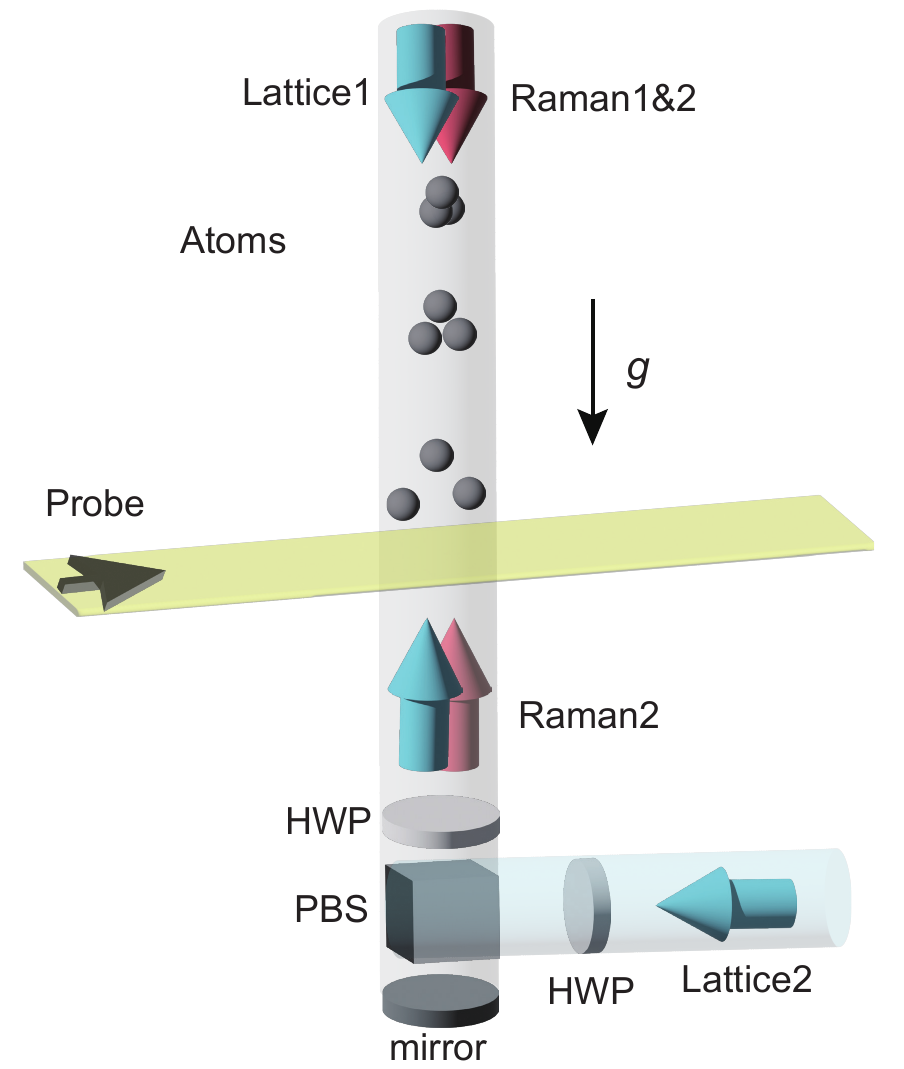}
\caption{\textbf{Experimental setup.} Cold rubidium atoms free fall from a moving optical lattice (cyan arrows) along the gravity direction. The moving lattice consists of two counter-propagating beams with a tunable frequency difference. Raman pulses (pink arrows) are used to split and recombine atomic clouds. Two Raman beams are shined from the top, rotated by a half-wave plate (HWP) and retroreflected by the mirror at the bottom. The upper Raman beams has two linearly polarized light and they are orthogonal. One of them is filtered out by the PBS and the other one is reflected upwards. The probe beam is shined perpendicular to the gravity direction for detecting the interference pattern. \label{figS1:Setup}}
\end{figure}

\begin{figure}[h]
\centering
\includegraphics[width=0.48\textwidth]{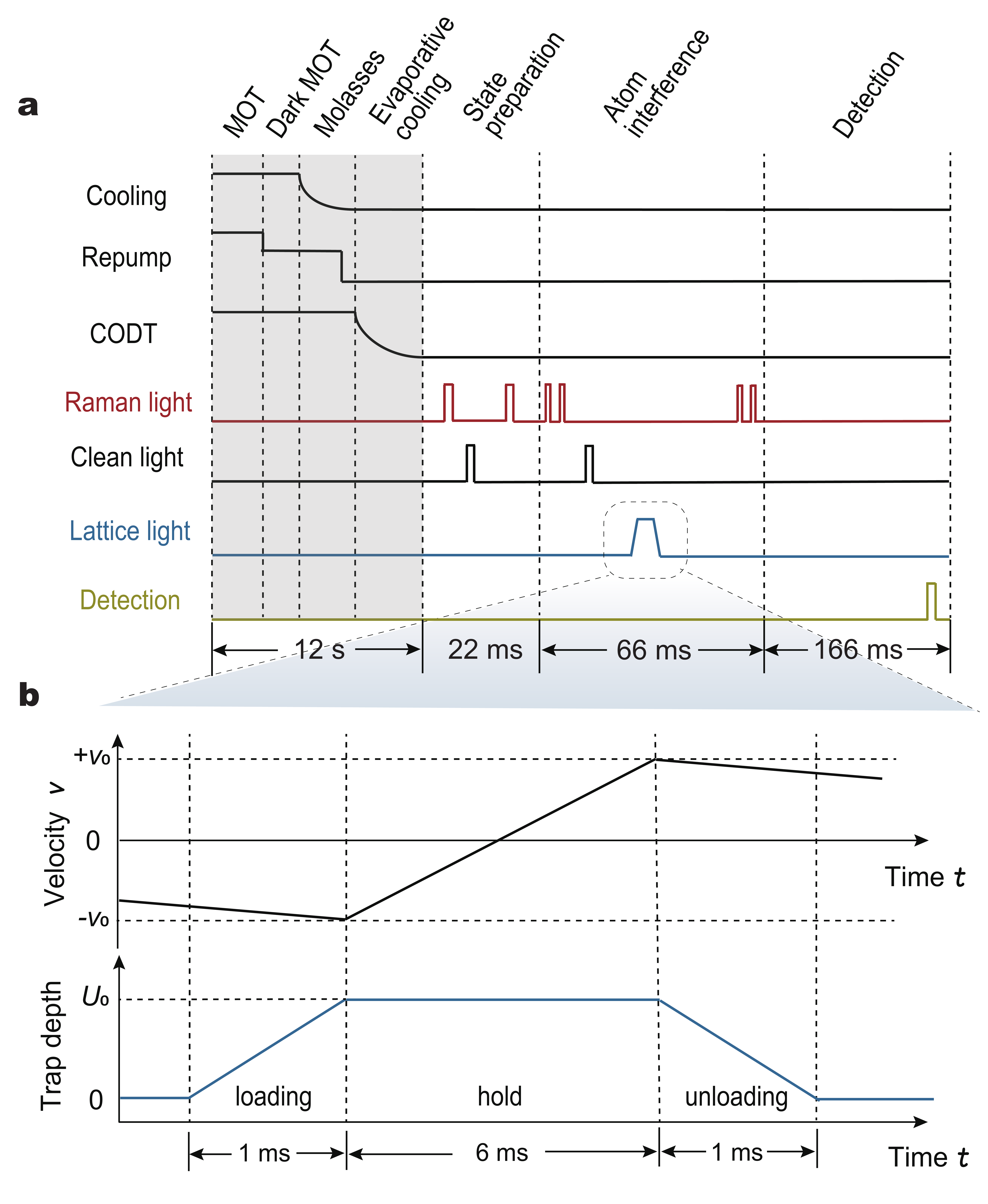}
\caption{\textbf{The experimental sequence.} (\textbf{a}) The sequence consists of the preparation of atoms (shaded grey area) and the atom interferometry (white area). (\textbf{b}) A zoomed-in sequence of the moving lattice. The upper and lower figures show the velocity change of the moving lattice and the trap depth. The velocity is tuned by the frequency difference during lattice loading and unloading to ensure that the atoms remain relatively stationary with respect to the moving lattice. 
\label{figS:sequence}}
\end{figure}

The wavelength of the moving optical lattice is $\lambda_{lat} = 780\,$nm, blue-detuned by $45\,$GHz from the resonant transition, and the lattice constant is $d=\lambda_{lat}/2 = 390\,$nm. Atoms are loaded at the quasi-momentum $q=0$. The frequency difference between the two lattice beams is optimized during lattice loading so that the atoms remain relatively stationary with the lattice. The time of lattice loading is $1\,$ms. 
The hold time in the optical lattice is $6\,$ms, during which atoms undergo 59 Bloch oscillations. After the lattice unloading within 1 ms, atoms are released from the lattice. Fig.~\ref{figS1:Setup} and Fig.~\ref{figS:sequence} show the experimental setup and the sequence respectively.

\paragraph{\textbf{Atom Interferometry.}}

Ramsey-Bord\'e consists of two pairs of Raman pulses to split and recombine the atomic cloud, shown in Fig.\ref{fig:Schematic}. The first $\pi/2$ pulse induces a $50\%$ probability of the stimulated Raman transition between the $\ket{F=1}$ state and the $\ket{F=2}$ state of $^{87}$Rb atoms \cite{kasevichAtomic1991}. The transferred atoms in the $\ket{F=2}$ state obtain the extra momenta induced by the Raman transition and move upwards with respect to the original $\ket{F=1}$ state.

After the first pair of Raman pulses, the two arms of the interferometer are in the same sate and momentum, $\ket{F=1, p=p_0}$, while atoms in the unwanted arms, $\ket{F=2, p=p_0+\Delta p}$ are removed by a clean beam on resonance of the $|F=2\rangle\rightarrow|F'=3\rangle$ transition. 
Atoms in two arms free fall and remain relatively stationary with respect to the lattice frame along the gravity direction during lattice loading and unloading. The frequency difference between the two lattice beams is scanned at $25.1\,\rm{MHz/s}$ to compensate the chirp rate of the Doppler frequency shift caused by the gravity. Atoms gain momentum transfer during Bloch oscillations in the moving lattice~\cite{peikBloch1997} and are coherently kicked upwards by the lattice. This kick effect can be used to fold the trajectory and acquire a longer evolution time for the interferometer to achieve higher sensitivity in a limited space~\cite{andiaCompact2013}.  

In the experiment, the Raman coupling is detuned by $\delta$ from the resonance in the second pair of Raman pulses. We scan the $\delta$ in the experiment and measure the transition probability $\mathcal{P}$ in Fig.~\ref{fig:Ramsey}a in the main text.

\begin{figure}[h]
\centering
\includegraphics[width=0.48\textwidth]{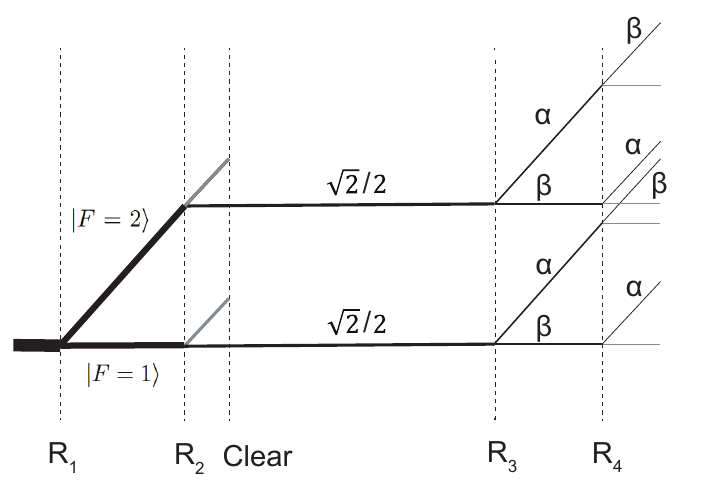}
\caption{\textbf{The weight change of atomic wave functions along the interferometer paths.}
Weights of wave functions are marked after Raman pulses along the interferometer paths. $R_1$-$R_4$ and Clear indicate the Raman pulses and the cleaning pulses of removing the $\ket{F=2}$ state, respectively. Grey lines indicate paths which are not involved in the atom interfereometer. 
\label{figS:theory}}
\end{figure}

\paragraph{\textbf{Interference fringe.}} We simulate fringes of the atom interferometer using a semi-classical approach based on the phase and weight of the wave functions along interferometer paths shown in Fig.~\ref{figS:theory}. The Raman transfer coefficient,
$\alpha=(\Omega/\sqrt{\Omega^2+\delta^2})\sin{\left(t_\text{L}\sqrt{\Omega^2+\delta^2}/2\right)}$,
where $\Omega$ is the Rabi frequency and $t_\text{L}$ is the pulse duration, and the rest state is thus $\beta=\sqrt{1-\alpha^2}$. 

The initial weight of each wave function is equally split from the same cloud and thus has a factor of $\sqrt{2}/2$. After two Raman pulses, two paths of the $\ket{F=2}$ state at the output port of the interferometer read,
\begin{equation}
\begin{aligned}
\label{fringe wave function}
&\psi_1=\frac{\sqrt{2}}{2}\alpha \beta e^{i\phi_c} \\
&\psi_2=\frac{\sqrt{2}}{2} \alpha \beta e^{i(\phi_c+\phi_d)}
\end{aligned}
\end{equation}
where $\phi_c$ is the common phase for the two wave functions, and the differential phase  $\phi_d=T_R  \delta+\phi_0$ with $\phi_0$ being the offset.
The probability  of the $\ket{F=2}$ state is, 
\begin{equation}
\label{diff phase}
\mathcal{P}=\lvert\psi_1+\psi_2\rvert^2+\frac{\sqrt{2}}{2} \alpha \beta  \times 2
\end{equation}
where the last term from the other two diverging paths of the $\ket{F=2}$ state only contributes an offset to the fringe, as it exceeds the coherence length of $\psi_1$ and $\psi_2$ (see Fig.~\ref{figS:theory}).
The above interference fringe $\mathcal{P}$ is plotted as the function of $\delta$ in Fig.~\ref{fig:Ramsey}(a) in the main text, where a fudge factor of 0.49 is multiplied  and an offset of 0.07 to count the imperfect Raman transition and detection in the simulation.

\paragraph{\textbf{Phase of the interferometer.}} The accumulated phase of the atom interferometer $\Delta \varPhi=\Delta \varPhi_{\text{Prop}}+\Delta \varPhi_{\text{Laser}}$, consists of the propagation phase $\Delta \varPhi_{\text{Prop}}$ and the laser phase $\Delta \varPhi_{\text{Laser}}$. $\Delta \varPhi_{\text{Prop}}$ is proportional to the integration of Lagrangian, contributed by the kinetic energy and the potential energy,

\begin{equation}
\begin{aligned}
\label{prop phase}
\Delta \varPhi_{\text{Prop}}=\frac{1}{\hbar}\int_{t_0}^{t_\text{d}} \left(\frac{1}{2}mv^2_\text{up}-\frac{1}{2}mv^2_\text{low}\right) \, dt \\
-\frac{mg}{\hbar}\int_{t_0}^{t_\text{d}}\left(z_\text{up}-z_\text{low}\right) \, dt
\end{aligned}
\end{equation}
where subscripts "up" and "low" indicate upper and lower arms respectively, $t_0$ and $t_\text{d}$ is the time of the first Raman pulse and  the detection time, respectively. The laser phase $\Delta \varPhi_{\text{Laser}}$ is determined by the time and the position of the laser beam,

\begin{equation}
\begin{aligned}
\label{laser phase}
\Delta \varPhi_{\text{Laser}}=\varPhi_1-\varPhi_2-\varPhi_3+\varPhi_4 ,\\
\varPhi_i=k_\text{eff}z_i-\int_{t_0}^{t_i} \omega_R \,dt
\end{aligned}
\end{equation}
where $z_i$ denotes the position at which atoms interact with the \emph{i}-th Raman pulse, and $\omega_R$ is the Raman frequency, which is chirped to compensate the Doppler frequency shift of free-falling atoms. The +/- sign before $\varPhi_i$ indicates the direction of the Raman transition from the state $\ket{F=1}$ to $\ket{F=2}$  or $\ket{F=2}$ to $\ket{F=1}$, respectively. Since $\Delta \varPhi_{\text{Laser}}$ is the differential phase between two arms, the phase accumulation from the optical lattice is cancelled out, whereas the one from four Raman pulses remains.

The phase difference of interferometers induces the differential phase between different interferometers. The phase difference for two interferometers in Fig.~\ref{fig:Diff}(a) are $\Delta \varPhi_\text{out}$ and $\Delta \varPhi_\text{in}$, giving rise to a beat when scanning the differential phase,

\begin{equation}
\label{diff phase}
\Delta \varPhi_\text{diff}=\Delta \varPhi_\text{out}-\Delta \varPhi_\text{in}=\frac{mgd}{\hbar}\tau
\end{equation}
where the inhomogeneity of the gravity is ignored. Thus the space-time area is a rectangle enclosed by the two arms in Fig.~\ref{fig:Diff}(a) in the main text.

\paragraph{\textbf{Contrast of the interference.}}
The fringe contrast is determined by the coherence of the system and the atomic distribution. The coherence length along the lattice direction is estimated using the interference of two arms~\cite{Longitudinal}. We assume the wave function in one arm as the integral of plane waves over the momentum $p$,

\begin{equation}
\label{wave function with p}
\psi(z)=\frac{1}{(\sqrt{2\pi}\sigma_p)^{1/2}}\int{e^{-\frac{p^2}{4\sigma^2_p}}}e^{\frac{i}{\hbar}(p z-\frac{p^2}{2 m} t)}\,dp
\end{equation}
where the standard deviation $\sigma_p$ of the momentum distribution  along the gravity direction is estimated by the equipartition theorem $\sigma_{p}^2/m^2=k_B T^2$. The other arm of the interferometer can be written as $\psi(z-\delta L) e^{i \phi_m}$, where $\delta L$ is the space between two arms, and $\phi_m$ is the accumulated phase from different paths and laser pulses. Therefore, the overall probability is the integration over the $z$ direction,
\begin{equation}
\begin{aligned}
\label{auto correlation}
P_{\delta L} &=\frac{1}{4}\int\lvert\psi(z)+\psi({z-\delta L})e^{i \phi_m}\rvert^2\,dz \\
&=\frac{1}{2}+\frac{1}{2\sqrt{2\pi}\sigma_p}\int{e^{-\frac{p^2}{2\sigma^2_p}}}\cos{\left(\frac{p}{\hbar}\delta L+\phi_m\right)}\,dp
\end{aligned}
\end{equation}

The fringe contrast $\mathcal{C}$ reads,

\begin{equation}
\begin{aligned}
\label{coherence length}
\mathcal{C}(P_{\delta L})&=\frac{1}{2\sqrt{2\pi}\sigma_p}\int{e^{-\frac{p^2}{2\sigma^2_p}}}\cos{\left(\frac{p}{\hbar}\delta L \right)}\,dp \\
&\approx \mathcal{C}_0 + A e^{-\frac{2\delta L^2}{\xi^2}}
\end{aligned}
\end{equation}
where $\mathcal{C}_0$ and $A$ are the offset and amplitude respectively. The extracted coherence length is $\xi=2\hbar/\sigma_p$ when $\sigma_p$ approaches zero
\cite{parazzoliObservation2012, Longitudinal}.

In our experiment, the time of lattice loading and unloading is chosen to be relatively short to induce the lattice modulation to the atomic cloud. For a deep lattice, $V_{lat}=40E_r$, the overlap of maximally localized Wannier functions  $w(z)$ between adjacent sites is negligible. Thus the system wave function $W(z)$ is the sum of Wannier functions, 
\begin{equation}
W(z)=\frac{1}{(\sqrt{2\pi}N)^{1/2}}e^{-\frac{z^2}{4N^2}}\sum_{z'=-4N}^{4N}w\left(\frac{z+z'}{d'/d}\right)
\end{equation} 
where $N$ denotes the Gaussian envelope to normalize the modulated wave function $\psi(z)W(z)$ and is set to be $10 d$, that is  much smaller than the actual size (hundreds of times $d$). $d'$ denote the modulated spacing after the expansion at the detection time. If we assume the atomic cloud expands ballistically as
$\sigma(t)^2=\sigma^2_0+\sigma^2_p t^2/m^2$ with the initial spread $\sigma_0$. We have
$d'/d=\sigma(T_0+T_{R1}+T_1+T_2+T_{R2}+T_3)/\sigma(T_0+T_{R1}+T_1)$,
where $T_0$ denotes the time interval from the cloud release to the first Raman pulse, and $T_i$ and $T_{Ri}$ denote the intervals between the pulse sequences in Fig.\ref{fig:Schematic} in the main text. This contributes to the outward shift of the peaks in Fig.~\ref{fig:Ramsey}(b-d) in the main text.

Substitute $\psi(z)$ with the modulated wave function $\psi(z)W(z)$, the modulated contrast  $\mathcal{C}'(P_{\delta L})$ can be derived as follows, 

\begin{equation}
\label{Wannier multiple}
\mathcal{C}'(P_{\delta L})=\mathcal{C}(P_{\delta L})\times\int{W(z)W(z-\delta L)}\,dz
\end{equation}
For the simulation in Fig.~\ref{fig:Ramsey}(b-d), the initial size is 380$\,d$, and the parameter $\sigma_p$ is chosen to fit the experimental data.

\begin{figure*}
\centering
\includegraphics[width=\textwidth]{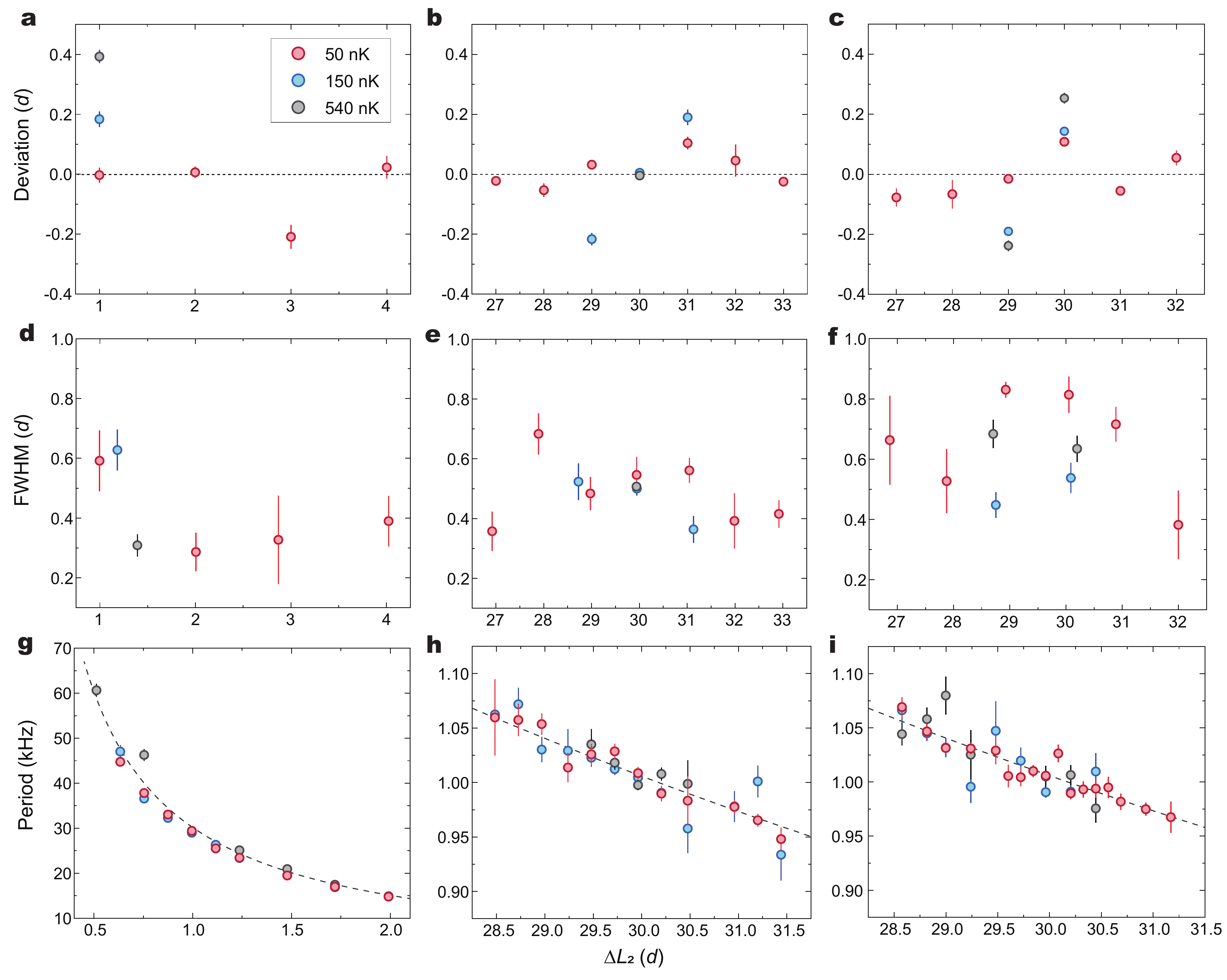}
\caption{\textbf{Characteristics of multiple peaks.} 
Position deviation, full width at half maximum (FWHM) and fringe periods of multiple peaks for different temperatures are extracted by fitting voigt curves to the data in Fig.\ref{fig:Ramsey} in the main text. (\textbf{a, d, g}), (\textbf{b, e, h}) and (\textbf{c, f, i}) are the intraference, the integer lattice loading with $\Delta L_1=30d$, and the half-integer lattice loading with $\Delta L_1=29.5d$ respectively.
(\textbf{a-c}) show the shift of peaks from the lattice position and dashed lines indicate zero. The shift increases with increasing the temperature. (\textbf{d-f}) are FWHMs for different temperatures. (\textbf{g-i}) are the periods of Ramsey fringes in good agreement with theoretical calculations (dashed lines). 
\label{figS:peaks}}
\end{figure*}

\paragraph{\textbf{Analysis of the peaks.}} 
Fig.~\ref{fig:Ramsey} in the main text shows that the peak position of the fringe contrast mismatches with the optical lattice sites, particularly when the coherence length becomes short. We fit a voigt curve to the data, and extract the shift and the width of peaks in Fig.~\ref{figS:peaks}. The error bar indicates the standard deviation of the fitting.

Fig.~\ref{figS:peaks}(a-c) show the temperature-dependent deviation. When the temperature of atoms is cold ($T=50\,$nk), peaks match the lattice sites, whereas the shift becomes larger with increasing temperature. The widths of peaks for different temperatures are extracted in Fig.~\ref{figS:peaks}(d-f).
Fig.~\ref{figS:peaks}(g-i) show the period of Ramsey spectroscopy which is extracted by fitting a sinusoidal curve to Ramsey fringes. The period of Ramsey fringes is inversely proportional to $\Delta L_2$. Fig.~\ref{figS:peaks}(g-i) are the collapsed experimental data on the theoretical curves, and show that the period is independent on the temperature.

\end{document}